\documentclass[a4paper,11pt]{article}
\usepackage{pos}
\newcommand{\pd}[1]{\, \partial #1 \,}

\newcommand{\g}{\ensuremath{\gamma}}

\title{ExHaLe-jet: Modeling blazar jets with an extended hadro-leptonic radiation code}
\ShortTitle{ExHaLe-jet}

\author*[a,b]{Michael Zacharias}
\author[c]{Anita Reimer}
\author[a]{Catherine Boisson}
\author[a]{Andreas Zech}

\affiliation[a]{Laboratoire Univers et Th\'eories, Observatoire de Paris, Université PSL, CNRS, Universit\'e de Paris, 92190 Meudon, France}

\affiliation[b]{Centre for Space Science, North-West University, Potchefstroom, 2520, South Africa}

\affiliation[c]{Institut f\"ur Astro- und Teilchenphysik, Leopold-Franzens-Universit\"at Innsbruck, A-6020 Innsbruck, Austria}

\emailAdd{michael.zacharias@obspm.fr}
\emailAdd{mzacharias.phys@gmail.com}

\abstract{Blazars emit across all electromagnetic wavelengths. While the so-called one-zone model has described well both quiescent and flaring states, it cannot explain the radio emission and fails in more complex data sets, such as AP Librae. In order to self-consistently describe the entire electromagnetic spectrum emitted by the jet, extended radiation models are necessary. Notably, kinetic descriptions of extended jets can provide the temporal and spatial evolution of the particle species and the full electromagnetic output. Here, we present the initial results of a newly developed hadro-leptonic extended-jet code: ExHaLe-jet. As protons take much longer than electrons to lose their energy, they can transport energy over much larger distances than electrons and are therefore essential for the energy transport in the jet. Furthermore, protons induce injection of additional pairs through pion and Bethe-Heitler pair production, which can explain a dominant leptonic radiation signal while still producing neutrinos. In this talk, we discuss the differences between leptonic and hadronic dominated SED solutions, the SED shapes, evolution along the jet flow, and jet powers. We also highlight the important role of external photon fields, such as the accretion disk and the BLR.}

\FullConference{%
  7th Heidelberg International Symposium on High-Energy Gamma-Ray Astronomy (Gamma2022)\\
  4-8 July 2022\\
  Barcelona, Spain\\}

\begin{document}
\maketitle

\section{Introduction}
In this proceedings paper, we provide a short overview of a recently developed extended hadro-leptonic jet code -- \textit{ExHaLe-jet} \cite{zacharias+22}. We also show four applications and discuss their impact.

The so-called one-zone model of blazars -- where a single, spherical, typically homogeneous emission region is responsible for most of the radiative output -- is widely used to reproduce flaring events. However, its validity to describe the quiescent state of a jet is at least questionable. In fact, it has been shown in almost all energy bands that jets of active galaxies are capable of producing radiation in these energy bands on various distance scales up to several kpc from the black hole \cite{hess20}. Hence, jets are capable of producing highly relativistic particles at all length scales. In order to successfully model these various acceleration sites and the connection between them, kinetic radiation codes are required that model the particle evolution and the radiation within the jet flow from its base to its termination \cite{pc13,lucchini+19}.

While most models use a leptonic scenario, the potential association of neutrinos with blazar jets demands the presence of relativistic protons in the jet frame. It is therefore important to co-evolve the protons and the related secondary particles alongside the electrons. While MHD, RMHD and GRMHD codes have improved (and continue to do so) to model jets on vast scales \cite{fichet+21}, the efficient calculation of the various kinds of radiation processes \cite{cerruti20} is best done with kinetic models. In such models, the Fokker-Planck equation governing the particle distribution under influences of injection, acceleration, cooling and other (catastrophic) losses, is solved along the jet flow by cutting the jet into numerous slices and imposing a fixed jet geometry and bulk-flow evolution \cite{pc13,lucchini+19}.

\textit{ExHaLe-jet} follows that description. It describes the evolution of all charged particle species and incorporates the corresponding interactions with the ambient photon fields. In addition to the internally produced photon fields (such as synchrotron), external photons from the accretion disk (AD), the broad-line region (BLR), and the dusty torus (DT) are also considered. As shown below, the external fields play a crucial role in the evolution of the jet.

\section{Code description}
\begin{figure}
\centering
\includegraphics[width=1.00\textwidth]{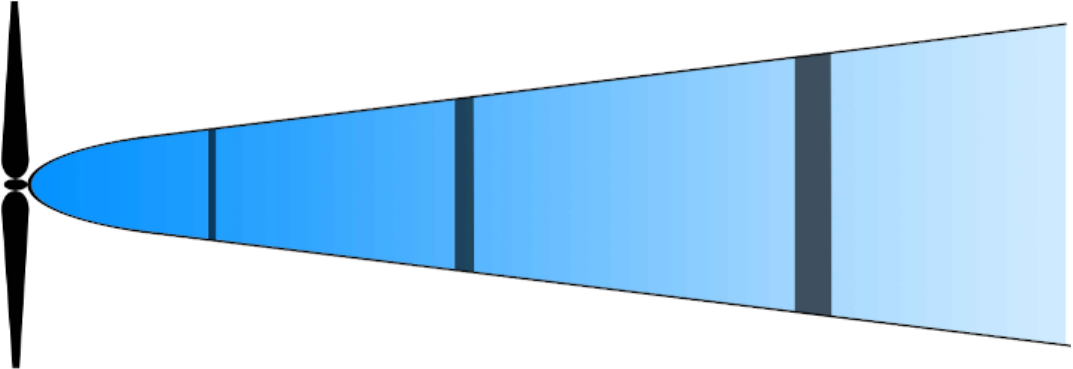}
\caption{Sketch of the model. The jet is cut into numerous slices (the dark regions exemplify these), where the Fokker-Planck equation is solved. Figure courtesy of Jonathan Heil.}
\label{fig:sketch}
\end{figure}
We employ a jet geometry with an initial parabolic bulk-flow acceleration region for $z<z_{acc}$ followed by a conical coasting region. The bulk Lorentz factor $\Gamma_b$ and the jet radius $R$ thus evolve as a function of jet distance $z$:
\begin{align}
	\Gamma_b(z) &\propto \sqrt{z}\qquad\qquad\qquad \mbox{for}\ z\leq z_{acc} \label{eq:Gb1} \\
	\Gamma_b(z) &= \Gamma_{b,{\rm max}} = \mbox{const}\quad\ \, \mbox{for}\ z> z_{acc} \label{eq:Gb2} \\
	R(z) &\propto \tan{\left[0.26/\Gamma_b(z)\right]}. \label{eq:R}
\end{align}
In order to calculate the radiative output of the entire jet, we cut it into numerous slices in a logarithmically-spaced grid along the $z$-axis. A sketch is shown in Fig.~\ref{fig:sketch}.

In each slice, we solve the time-dependent Fokker-Planck equation of the particle distribution for protons, charged pions, muons and electrons (including positrons). Additionally, the radiation transport equation is solved allowing for the direct feedback of the particle and photon interactions. The Fokker-Planck equation for the particle distribution of species $i$ is

\begin{align}
     \frac{\pd{n_i(\chi,t)}}{\pd{t}} &= \frac{\pd{}}{\pd{\chi}} \left[ \frac{\chi^2}{(a+2)t_{\rm acc}} \frac{\pd{n_i(\chi,t)}}{\pd{\chi}} \right] \nonumber \\
	 &\quad - \frac{\pd{}}{\pd{\chi}} \left( \dot{\chi}_i n_i(\chi,t) \right) + Q_i(\chi,t)
	 - \frac{n_i(\chi,t)}{t_{\rm esc}} - \frac{n_i(\chi, t)}{\gamma t^{\ast}_{i,{\rm decay}}}
	 \label{eq:fpgen}.
\end{align}
The distributions $n_i$ are given as a function of normalized momentum $\chi=\gamma\beta$, with the particle Lorentz factor $\gamma$ and its corresponding speed $\beta$ normalized to the speed of light. This ensures stability of the numerical scheme. The first term on the right-hand-side represents Fermi-II acceleration using hard-sphere scattering with the ratio $a$ of shock to Alfv\`{e}n speed. The second term marks continuous momentum gains through Fermi-I acceleration and losses. Continuous losses depend on the particle species and include synchrotron, adiabatic, Bethe-Heitler, pion production, and inverse-Compton processes. The third term in Eq.~(\ref{eq:fpgen}) marks the injection term, while the forth term represents the catastrophic escape of particles from one slice to the next. The last term is the decay term for unstable particles. In the current version of the code, neutrons are not explicitly considered.

Particles are assumed to escape only in the downstream direction but not through the side. It is described by the escape time scale $t_{\rm esc}=\eta_{\rm esc}\Delta z/c$ with the length of a slice $\Delta z$, and the multiple $\eta_{\rm esc}>1$ mimicking advection. The acceleration time scale is given as a multiple of the escape time scale: $t_{\rm acc} = \eta_{\rm acc}t_{\rm esc}$. It merely represents the reacceleration of particles in the slice, and does not provide ``first-principle'' acceleration, which typically requires a much smaller zone \cite{dmytriiev+21}. The initial acceleration is mimicked through the primary injection term $Q_i$, which for protons and electrons takes the form of a power-law between minimum and maximum Lorentz factors, $\gamma_{\rm i,min}$ and $\gamma_{\rm i,max}$, respectively, with spectral index $s_i$. Primary protons and electrons are injected at the base of the jet, and their normalization is then evolved along the jet according to the geometry of the jet. The injection of pions and muons is calculated from the respective interactions and decays using the template approach of \cite{huemmer+10}. Secondary electrons are injected from muon decay, Bethe-Heitler pair production, and \g-\g\ pair production. These secondary electrons are included in the jet flow and passed on to subsequent slices.

For particle-photon and photon-photon interactions we consider all available photons. These include external photon sources, namely the AD, the BLR, and the DT. The magnetic field is evolved following the relativistic Bernoulli equation \citep{Zdziarski+15}.

All details of the code are given in \cite{zacharias+22}.

\section{Results}
\begin{table*}
\caption{Parameters for the models in Figs.~\ref{fig:model01} and~\ref{fig:model02}. BLR and DT luminosities are assumed to be 10\% of the AD luminosity.
}
\begin{tabular}{lcccc} 
Simulation / Figure & 01A & 01B & 02A & 02B \\
Main $\gamma$-ray production process	& EC & EC & P-Syn & SSC \\
\hline
Jet length & $100\,$pc & $100\,$pc & $100\,$pc & $100\,$pc \\
Acceleration zone length & $1\,$pc & $1\,$pc & $1\,$pc & $0.1\,$pc \\
Disk Eddington ratio & $0.1$ & $0.01$ & --- & --- \\
Max. Doppler factor & $30$ & $30$ & $50$ & $30$ \\
Initial magnetic field & $50\,$G & $50\,$G & $70\,$G & $30\,$G \\
Injection particle power & $10^{-5}\,L_{\rm edd}$ & $10^{-5}\,L_{\rm edd}$ & $2\times10^{-4}\,L_{\rm edd}$ & $2\times10^{-6}\,L_{\rm edd}$ \\
Initial proton to electron ratio & $1$ & $1$ & $1$ & $10^{-10}$ \\
Proton $\gamma_{\rm min}$ / $\gamma_{\rm max}$ & $2$ / $2\times 10^{8}$ & $2$ / $2\times 10^{8}$ & $2$ / $2\times 10^{9}$ & $2$ / $2\times 10^{2}$ \\
Electron $\gamma_{\rm min}$ / $\gamma_{\rm max}$ & $100$ / $2\times 10^{4}$ & $100$ / $2\times 10^{4}$ & $100$ / $2\times 10^{4}$ & $10^{5}$ / $2\times 10^{6}$ \\
P \& e spectral index & $2.5$ & $2.5$ & $2.0$ & $2.8$ \\
\hline
Total jet power & $2\times 10^{-3}\,L_{\rm edd}$ & $2\times 10^{-3}\,L_{\rm edd}$ & $0.03\,L_{\rm edd}$ & $2\times 10^{-4}\,L_{\rm edd}$ \\
\end{tabular}
\label{tab:model}
\end{table*}
\begin{figure}
\centering
\includegraphics[width=1.00\textwidth]{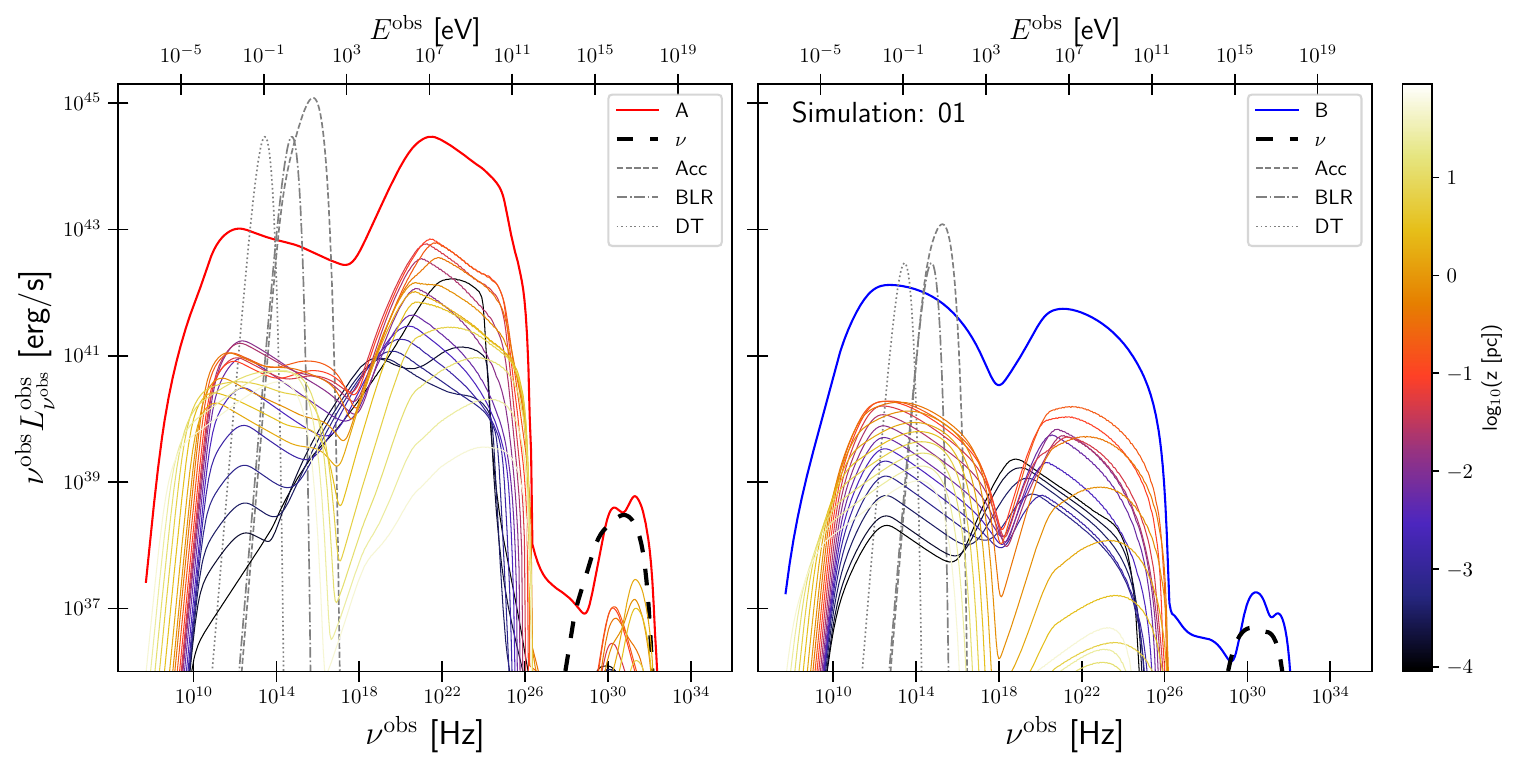}
\caption{Total luminosity (thick solid lines) as well as the evolution as a function of jet distance $z$ (color code) for strong external fields (left) and weak external fields (right). The thin gray lines mark the external fields as labeled, while the thick black dashed line marks the total muon-neutrino spectrum.}
\label{fig:model01}
\end{figure}
\begin{figure}
\centering
\includegraphics[width=1.00\textwidth]{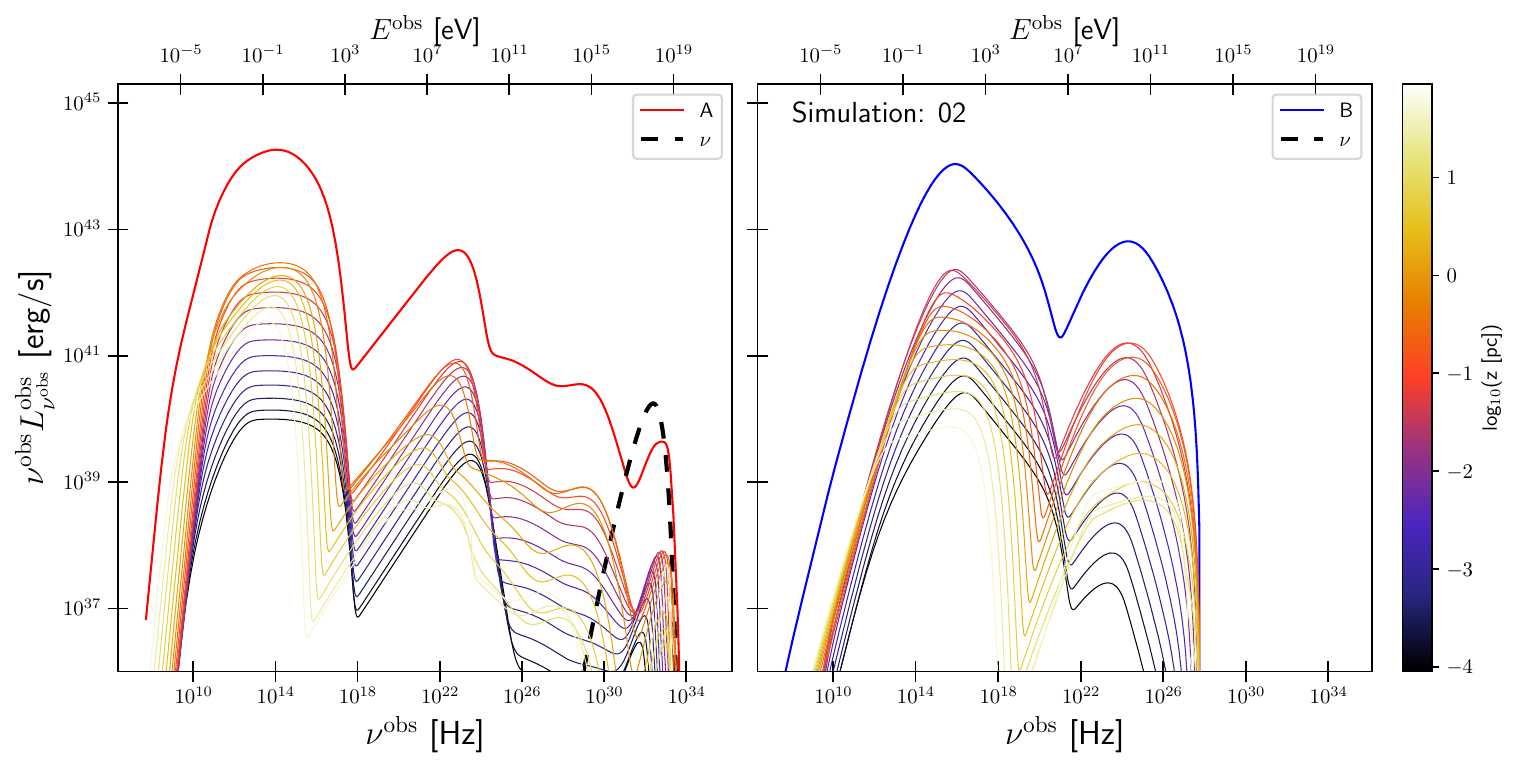}
\caption{Same as Fig.~\ref{fig:model01}, but without external photon fields and for a solution where the \g\ rays are dominated by proton-synchrotron (left) or SSC (right).}
\label{fig:model02}
\end{figure}
In Figs.~\ref{fig:model01} and~\ref{fig:model02} we display 4 sets of simulations, while the corresponding parameters are given in Tab.~\ref{tab:model}. The simulations in Fig.~\ref{fig:model01} show the influence of the external fields. Especially a strong AD induces significant pair cascades in the initial slices of the jet raising substantially the pair content. In turn, the radiation spectra are electron-dominated with the \g\ rays being produces predominantly through inverse-Compton processes involving BLR and DT photons. This also holds for weaker external fields (Fig.~\ref{fig:model01}, right), but the Compton dominance is less than unity in this case. The additional hump at frequencies $\sim 10^{30}\,$Hz stems from the neutral-pion decay. It is a coincidence that the neutrino flux (thick black dashed line) looks comparable to the neutral pion bump in these simulations. The neutral pion flux is strongly absorbed, and thus cannot serve as a proxy for the charged pion production, which is responsible for the neutrino production through their decays.

In order to derive proton-synchrotron and SSC dominated (in the \g-ray domain) solutions as displayed in Fig.~\ref{fig:model02}, we switch off the external photon fields. For the proton-synchrotron solution, we require a higher injection particle power, a higher initial magnetic field, and to allow for highly relativistic protons compared to the previous solutions. Next to the significant proton-synchrotron emission, proton-photon interactions are also increased, as indicated by the increased neutrino flux compared to Fig.~\ref{fig:model01}. However, this also injects significant numbers of pairs increasing the electron-synchrotron flux in the optical domain, as well as inducing the plateau beyond TeV \g-ray energies, which is synchrotron emission of secondary pairs.

For the SSC solution (Fig.~\ref{fig:model02}, right) we choose a shorter bulk-flow acceleration region. As the jet reaches its maximum speed (and thus maximum beaming) at a smaller distance from the black hole, the jet is more compact implying a higher density of particles and radiation. In turn, SSC is enhanced. One can see by the distance evolution that the SSC then drops rapidly -- and much faster than the synchrotron emission -- in the conical section of the jet. Without the external fields and the relativistic protons, no meaningful number of secondaries is injected in this case.

In all cases, the total jet power (incl. magnetic field, all particles, and the radiation) remains below the Eddington luminosity (Tab.~\ref{tab:model}). This is by design, as the Bernoulli equation only allows for a maximum power of particles depending on the initial magnetic field \cite{zacharias+22}. Within the described circumstances and setups, it is difficult to achieve the higher powers required for a detectable amount of neutrinos.

While \textit{ExHaLe-jet} is not yet capable of reproducing the various acceleration sites, it is a first step in this direction. Implementing discrete acceleration regions is the next planned step. This will be followed-up by exploting the time-dependency that is already available in Eq.~(\ref{eq:fpgen}). This will enable to study the evolution of blazar light curves in a more realistic way than efforts based on the one-zone model.

\section*{Acknowledgement}
MZ acknowledges postdoctoral financial support from LUTH, Observatoire de Paris.
AR acknowledges financial support from the Austrian Science Fund (FWF) under grant agreement number I 4144-N27.
Simulations for this paper have been performed on the TAU-cluster of the Centre for Space Research at North-West University, Potchesftroom, South Africa.


\end{document}